\documentclass[12pt]{article}
\usepackage{amsmath,amsthm,amsfonts,amssymb}
\usepackage{mathrsfs}

\begin{document}

\title{Gauge model of the fifth force}

\author{G. A. Sardanashvily,\and E. G. Timoshenko\thanks{Translated on 07.07.2024 by Edward.Timoshenko@ucd.ie from Russian: The Bulletin of Moscow State University, ISSN 2074-6636, UDC 530.12
Series 3, Physics and astronomy, Vol. 31, No 4, pp. 70-72 (1990).}
}

\date{
Division of Theoretical Physics, Department of Physics, Lomonosov Moscow State University, Russia, USSR
}

\maketitle

\begin{abstract}
The Lagrangian and equations of the gauge model of the fifth fundamental interaction are constructed and the corresponding corrections to the Newtonian potential are obtained.
\end{abstract}


\section{Introduction}
\label{sec:intro}

The hypothesis of the existence of the fifth fundamental interaction is now widely discussed in connection with certain ambiguous results of laboratory tests of the Newton's law of gravitation. The fifth force is broadly viewed as the set of hypothetical effects that could generate contributions of the ``Yukawa'' type to the Newtonian potential. We have proposed a model of such interaction using the gauge fields of the translational group [1,2]. Its peculiarity in comparison with other affine gauge theories consists in the construction of a deformation mapping of a space--time manifold and consideration of material fields on such a manifold.

Let $AX$ be the principal fibre bundle of affine frames with the structural affine group $A(4,R)$ over the space--time manifold $X^4$. We equip its total space $P$ with coordinates $\{x^{\mu},u^a,S^{\mu}_a\}$. Here $x^{\mu}$ are the coordinates in $X^4$, $u^a$ are the coordinates of the translation subgroup $T_4$, and $S^{\mu}_a$ are the coefficients of the linear connection, $\{St_a\}$ with respect to frame $\{\partial_{\mu}\}$, where $\{t_a\}$ is a fixed basis of $T_4$, and $S$ is an element of
$GL(4,R)$. Note that $\{x^{\mu},u^{\mu}=S^{\mu}_a u^a\}$ are the coordinates of the affine tangent fibre bundle $AT(X)$. Let a generalised affine connection be given in $AX$. Its connection form $\omega$ and the corresponding horizontal fields $t^h$ on $P$ are of the form
\[
\omega = (S^{-1})^a_\varepsilon (dS^\varepsilon_b + \Gamma^\varepsilon_{\mu\alpha}(x) S^\alpha_b dx^\mu) I^b_a + (du^a - B^a_\mu(x) dx^\mu) T_a,
\]
\begin{equation}
\tau^h = \tau^\mu(x) \left(\frac{\partial}{\partial x^\mu} + B^a_\mu(x) \frac{\partial}{\partial u^a} - \Gamma^\varepsilon_{\mu\alpha}(x) S^\alpha_b \frac{\partial}{\partial S^\varepsilon_b}\right),
\end{equation}
where $I^a_b,\ T_a$ are the generators of group 
$A(4,R)$, $\Gamma^{\varepsilon}_{\alpha\mu}$ are the coefficients of linear connection,
and $B^{\varepsilon}_{\mu}(x)=S^{\varepsilon}_a B^a_{\mu}(x)$ are gauge fields of the translation group. The physical interpretation of the
$B$ fields presents some difficulties because the material fields are sections of linear rather than of affine fibre bundle. The example of $B$ fields of 3-dimensional translations in dislocation theory [3] indicates a way out.

In order to give the meaning to the displacement vectors, let us consider the following mapping 
$\rho$ of the space $P$ onto the total space $Q$ onto the total sub--bundle of linear frames $Lx$ over $X^4$ at the points $S^{\varepsilon}_a u^a=u^{\varepsilon}(x)$:
\[
(x^\mu, u^a, S^\mu_a) \rightarrow (\gamma^\mu(x^\varepsilon; u^\varepsilon(x) - S^\varepsilon_a u^a; 1), 0, S^\mu_a) = (x^\mu, 0, S^\mu_a),
\]
where $\gamma(x; u; s)$ is a geodesic passing through $x$ in the direction $u$ and defined by the linear connection $\Gamma$, and $u(x)$ is some section of the fibre bundle $AT(X)$. 
The tangent  $\rho_*$ to $\rho$ mapping of a tangent  fibre bundle over $P$ to a tangent fibre bundle over
$Q$ transforms the fields (1) on $P$ into the fields
\begin{equation}
\tau^h_Q = \tau^\mu(x) \left(\delta^\nu_\mu + \mathcal{D}_\mu u^\nu(x)\right) \left[\frac{\partial}{\partial x^\nu} - \Gamma^\varepsilon_{\mu\alpha}(x) S^{\alpha}_b \frac{\partial}{\partial S^{\varepsilon}_b}\right]
\end{equation}
on $Q$, where the covariant derivative of the field and $u(x)$ has the form
\begin{equation}
\mathscr{D}_{\mu} u^{\varepsilon} (x) = \partial_{\mu} u^{\varepsilon} (x) + \Gamma^{\varepsilon}_{\mu\alpha} u^{\alpha} (x) - B^{\varepsilon}_{\mu} (x) = \sigma^{\varepsilon}_{\mu} (x) .
\end{equation}
Note that it is usual to take the mapping 
$\beta:\ (x^\mu, u^a, S^\mu_a) \rightarrow 
 (x^\mu, 0, S^\mu_a)$
of the space $P$ onto $Q$ at the points $S^{\varepsilon}_a u^a=u^{\varepsilon}(x)$
in affine gauge models.
At these points $\beta=\rho$, but  $\beta_*\not=\rho_*$. Comparing this to the case $B=0,\ u(x)=0$, the mappings $\rho,\ \rho_*$ can be interpreted as deformations of $X^4$ manifold. 
We should note that by a gauge transformation the translation field $u(x)$
can always be converted to zero. Therefore, 
only its covariant derivatives (3) are physically meaningful.

Let $\varphi$ be some tensor field on $X^4$ and $f_{\varphi}$ be the corresponding tensorial function on $Q$. We shall say that $\varphi$ is defined on a deformed manifold $X^4$ if its differentiation is defined in the form 
$(\mathcal{D}_\mu \varphi)(\tau)=(df_{\varphi})(\tau^h_Q)$,
where $\tau^h_Q$ is the horizontal (with respect to $\Gamma$) lift (2) of the $\tau=\tau^{\mu}\partial_{\mu}$ field. This means that in field theory the deformation of the manifold can be accounted for by replacing in the external differentiation operator $dx^{\mu}\mathcal{D}_\mu$
of the ordinary covariant derivatives $\mathcal{D}_\mu$ by $\widetilde{\mathcal{D}_\mu}=(\delta^\alpha_\mu+\sigma^\alpha_\mu)\mathcal{D}_\alpha=H^\alpha_\mu\mathcal{D}_\alpha$.

Let us write out, for example, the Lagrangian of a scalar field:
\[
L_{(\varphi)} = \frac{1}{2} [g^{\mu\nu} H^{\alpha}_{\mu} H^{\beta}_{\nu} \partial_{\alpha} \varphi \partial_{\beta} \varphi - m^2 \varphi^2]
\]
and the action functional and the equations of motion of the point mass:
\begin{equation}
S = - m \int [d_{\alpha\beta} H^{\alpha}_{\mu} H^{\beta}_{\nu} u^{\mu} u^{\nu}]^{1/2} ds,
\quad
\frac{du^{\mu}}{ds} + \widetilde{\Gamma}^{\mu}_{\alpha\beta} u^{\alpha} u^{\beta} = 0,
\end{equation}
where the values of $\widetilde{\Gamma}$ take the form of Christoffel symbols of the $H^{\alpha}_{\mu} H^{\beta}_{\nu} g_{\alpha\beta}$ ``metric'' but $ds$ is defined by the metric $g$. The Lagrangians of the gravitational and electromagnetic fields $L_{(g)}$ and $L_{(A)}$ are constructed in the usual way, but from the modified tensors  $H^{\varepsilon}_{\mu} H^{\beta}_{\nu} R^{ab}_{\varepsilon\beta}$ of the curvature and the strength  $H^{\alpha}_{\mu} H^{\beta}_{\nu} F_{\alpha\beta}$. 

The Lagrangian $L_{(\sigma)}$ of the gauge field of translations $B^\varepsilon_\mu$ itself cannot be constructed by the usual rules of Yang--Mills theory, since the Lie algebra of the affine group does not admit a nondegenerate invariant bilinear form. But it can be constructed from the values of $\sigma^\nu_\mu$ and $F^\alpha_{\mu\nu}=\mathcal{D}_{[\nu}\sigma^\alpha_{\mu]}$. Under the condition of positivity of the component $T^{00}_{(\sigma)}$ of the metric tensor energy--momentum of the field $\sigma$, it has the form $(a_1>0,\ a_2>0,\ \mu >0, \ \lambda < \mu/4,\ \sigma_{\mu\nu}=g_{\mu\alpha}\sigma^\alpha_\nu)$
\[
L_{(\sigma)} = \frac{1}{2} [a_1 F^{\mu}_{\nu \mu} F^{\nu \alpha}_{\alpha} + a_2 F_{\mu \nu \sigma} (F^{\mu \nu \sigma} - 2F^{\nu \mu \sigma}) - \mu \sigma^{\mu}_{\nu} \sigma^{\nu}_{\mu} + \lambda \sigma^{\mu}_{\mu} \sigma^{\nu}_{\nu}].
\]
The sources of the field $\sigma$ are: the shortened canonical energy--momentum tensor of material fields
\[
\frac{\delta L_{(m)}}{\delta \sigma^{\mu \nu}} = (H^{-1})_{\nu \beta} \mathscr{D}_{\mu} \varphi \frac{\partial L_{(m)}}{\partial \mathscr{D}_{\beta} \varphi} = (H^{-1})_{\nu \beta} (t^{\beta}_{(m)\mu} + \delta^{\beta}_{\mu} L_{(m)})
\]
the shortened metric energy-momentum tensor of the electromagnetic field $A$ and the curvature tensor $\kappa^{-1}H^\nu_\mu R^{\mu\gamma}_{\gamma\varepsilon}$
of the gravitational field. But instead of the latter, by using the 
Einstein's equations, we can substitute its expression through the metric energy--momentum tensors of matter and the field $\sigma$ itself. Then the equation for $\sigma$ takes the form of
\[
\frac{\delta L_{(\sigma)}}{\delta \sigma^{\mu\nu}} = \left[-\frac{\delta L_{(m)}}{\delta \sigma^{\mu\nu}} - (H^{-1})_{\nu\beta} \left(T^{\beta}_{(m)\mu} - \frac{1}{2} \delta^{\beta}_{\mu}T_{(m)}\right)\right] -
\]
\begin{equation}
-(H^{-1})_{\nu\mu} L_{(A)} + (H^{-1})_{\nu\beta} \left(T^{\beta}_{(\sigma)\mu} - \frac{1}{2} \delta^{\beta}_{\mu}T_{(\sigma)}\right).
\end{equation}

Let us restrict ourselves to the case of a weak field $\sigma$, neglecting in the left part of Eq. (5) the interaction of $\sigma$ with the gravitational field and torsion, and in the right--hand side by the members of $\sigma$. We obtain
\[
\frac{\delta L_{(\sigma)}}{\delta \sigma^{\mu\nu}} = a_1 (\eta_{\mu\nu}\partial^\varepsilon F_{\alpha\varepsilon}^{\alpha} - \partial_\mu F_{\alpha\nu}^{\alpha}) + 2a_2\partial^\varepsilon (F_{\mu\nu\varepsilon} + F_{\varepsilon\mu\nu} - F_{\nu\mu\varepsilon}) - \mu\sigma_{\mu\nu} + \lambda\eta_{\mu\nu}\sigma_{\alpha}^{\alpha}.
\]

Let us consider equation (5) in the void. Given the condition
\begin{equation}
\partial^\nu \frac{\delta L_{(\sigma)}}{\delta \sigma^{\mu\nu}} = -\mu \partial^\nu \sigma_{\mu\nu} + \lambda \partial_\mu \sigma = 0
\end{equation}
it splits into equations for the antisymmetric 
$\omega_{\mu\nu}=(1/2)\sigma_{[\mu\nu]}$ and the symmetric $\varepsilon_{\mu\nu}=(1/2)\sigma_{(\mu\nu)}$ $(e=\sigma^\alpha_\alpha)$
parts of the field $\sigma$:
\begin{equation}
4a_2\partial^{\varepsilon}(\omega_{\mu\varepsilon,\nu} + \omega_{\nu\mu,\varepsilon} - \omega_{\nu\varepsilon,\mu}) + 2a_1\omega_{\alpha[\nu,\mu]\alpha} - \mu\omega_{\mu\nu} = 0,
\end{equation}
\begin{equation}
a_1(\lambda/\mu - 1)[\eta_{\mu\nu} \,\Box \,e - e_{,\mu\nu}] + 2a_1\omega_{\alpha(\nu,\mu)\alpha} - \mu e_{\mu\nu} + \lambda\eta_{\mu\nu} e = 0.
\end{equation}
As a solution to equation (7) we choose $\omega=0$ and as for equation (8), by using convolution by indices $\mu,\nu$, we write it in the form of
\[
\Box \,e + m^2e = 0, \quad m^2 = \frac{\mu(\mu - 4\lambda)}{3a_1(\mu - \lambda)},
\]
\[
e_{\mu\nu} = \frac{\mu - \lambda}{3\mu}\left(\eta_{\mu\nu}e - m^{-2}\left(\frac{\mu - 4\lambda}{\mu - \lambda}\right)e_{,\mu\nu}\right).
\]
It admits plane--wave solutions.

In the presence of sources, expression (6) is not equal to zero and, in general, is not a gradient, so $\mu \not=0$. For example, in the case of a point mass $M$, the source in the right-hand side of equation (5) is $-(1/2)\eta_{\mu\nu}M\delta(r)$  and its stationary centrally symmetric solution has the form
\[
e_{rr} = -(\mu - \lambda)^{-1} \left( 3\lambda e_{00} + \frac{1}{2} M\delta(r) \right), \quad e_{\theta\theta} = -e_{00}r^2, \quad e_{\varphi\varphi} = -e_{00}r^2 \sin^2 \theta,
\]
\[
\frac{1}{r^2} \frac{\partial}{\partial r} r^2 \frac{\partial}{\partial r} e_{00} - m^2e_{00} = -\frac{1}{6} \frac{\mu}{a_1 (\mu - \lambda)} M\delta(r), \quad e_{00} = -\frac{\mu M}{6a_1 (\mu - \lambda) r} e^{-mr}. 
\]
Substituting this solution into the equation of motion of the test particle (4), we obtain the correction to the Newtonian gravitational potential:
\begin{equation}
\varphi' = \varphi + e_{00} = -\frac{M}{8\pi r} \left( \varkappa - \frac{\mu}{3a_1 (\mu - \lambda)} e^{-mr} \right).
\end{equation}

Note that the fifth interaction, which gives a contribution to the gravitational effects, must be as universal as gravitation. For this, in particular, its material source can be mass or other components of the energy--momentum tensor of matter and it must be described by a massive field, albeit with an unusually small mass. The above gauge model satisfies these conditions. Thus, the mass $m$ is expressed through the constants $\mu$ and $\lambda$ in the Lagrangian $L_{(\sigma)}$, which have the sense of the ``elasticity'' coefficients of the space--time manifold and can be considered small enough, but $\mu \not= 0,\ \lambda < \mu/4$.

$\ $
\\
\noindent
{\bf 
Manuscript (Brief Communications, Theoretical and Mathematical Physics)
received in the editorial office on: 29.09.89. 
}

\end{document}